\documentclass[reprint,twocolumn,
superscriptaddress,
 amsmath,amssymb,
 aps,
pra,
]{revtex4-1}
\usepackage{orcidlink}
\usepackage{graphicx}
\usepackage{dcolumn}
\usepackage{bm}

\usepackage{comment}
\usepackage{cancel}
\usepackage{cleveref}   
\usepackage{hyperref}   
\usepackage{nameref}    
\usepackage{braket}

\crefname{equation}{Eq.}{Eqs.}
\crefname{figure}{Fig.}{Figs.}
\crefname{table}{Table}{Tables}
\usepackage{tikz}
\usetikzlibrary{calc}
\newcommand{\tikzmark}[1]{\tikz[overlay,remember picture] \node (#1) {};}
\usetikzlibrary {arrows.meta}
\graphicspath{./pics}

\begin{document}

\title{Effective theory for strongly attractive one-dimensional fermions}

\author{Timothy G. Backert\orcidlink{0000-0002-5701-0280}}
\email{timothy\_george.backert@tu-darmstadt.de}
 \affiliation{Department of Physics, Technische Universität Darmstadt, 64289 Darmstadt, Germany}
 \author{Fabian Brauneis\orcidlink{0000-0002-3696-5256}}
\affiliation{Department of Physics, Technische Universität Darmstadt, 64289 Darmstadt, Germany}
\author{Matija Čufar\orcidlink{0000-0003-0734-2719}}
\affiliation{Dodd-Walls Centre for Photonic and Quantum Technologies, Auckland 0632, New Zealand}
\affiliation{Centre for Theoretical
Chemistry and Physics, New Zealand Institute for Advanced Study,
Massey University, Auckland 0745, New Zealand}
\author{Joachim Brand\orcidlink{0000-0001-7773-6292}}
\affiliation{Dodd-Walls Centre for Photonic and Quantum Technologies, Auckland 0632, New Zealand}
\affiliation{Centre for Theoretical
Chemistry and Physics, New Zealand Institute for Advanced Study,
Massey University, Auckland 0745, New Zealand}
\author{Hans-Werner Hammer\orcidlink{0000-0002-2318-0644}}
 \affiliation{Department of Physics, Technische Universität Darmstadt, 64289 Darmstadt, Germany}
\affiliation{Extreme Matter Institute EMMI and Helmholtz Forschungsakademie Hessen für FAIR (HFHF),
GSI Helmholtzzentrum für Schwerionenforschung GmbH, 64291 Darmstadt, Germany}
\author{Artem G. Volosniev\orcidlink{0000-0003-0393-5525}}%
\email{artem@phys.au.dk}
\affiliation{Department of Physics and Astronomy, Aarhus University, 8000 Aarhus C, Denmark}

\date{\today}

\begin{abstract}
We study a one-dimensional system of two-component fermions in the limit of strong attractive particle-particle interactions. First, we analyze scattering in the corresponding few-body problem, which is analytically solvable via Bethe ansatz. This allows us to engineer effective interactions between the system's effective degrees of freedom: fermions and bosonic dimers (tightly bound pairs of fermions). We argue that, although these interactions are strong, the resulting effective problem can be mapped onto a weakly interacting one, paving the way for the use of perturbation theory. This finding simplifies studies of many-fermion systems under confinement that are beyond reach of state-of-the-art numerical methods. We illustrate this statement by considering an impurity atom in a Fermi gas.
\end{abstract}

\maketitle

{\it Introduction.---} Understanding physical systems with strong electron-electron correlations, e.g., itinerant ferromagnets~\cite{Skomski2008}, requires exploration of
interacting spin-$\frac{1}{2}$ fermionic models. Typically, these models can only be solved approximately or within certain limits, motivating the search for exact solutions as foundational elements for further developments. One technique for generating such solutions is the mapping of a strongly interacting system onto a weakly-interacting one. The celebrated Bose-Fermi mapping~\cite{34girardeau1960relationship}, which maps strongly repulsive one-dimensional (1D) continuum models~\cite{MISTAKIDIS20231} onto non-interacting fermions, is an example relevant for this study. Here, we address a two-component system in the opposite limit of strong attractions, ultimately demonstrating that it too can be mapped onto a weakly-interacting problem.

In particular, we argue that strongly attractive systems can be linked to a weakly-interacting mixture of fermions and dimers (bound states between spin-up and spin-down fermions). Our focus is on trapped spin-imbalanced systems, whose solution is, as a rule, particularly difficult~\cite{BERGER20211}.
As a limit, this Letter also includes the known spin-balanced model -- a reference point for BEC-BCS studies~\cite{Astrakharchik2004,21fuchs2004exactly} as well as for corresponding few-body problems~\cite{Shamailov2016}.

{\it Framework.---} Our goal is to relate a trapped system of $N=N_\downarrow+N_\uparrow$  fermions of mass $m$,
\begin{equation}
        H=\sum_{i=1}^{N}\left[t_{x_i}+v_\text{ext}(x_i)\right]+g\sum_{i<j}^{N}\delta(x_i-x_j)\ ,\label{eq:GY Hamiltonian}
\end{equation}
in the limit of `large' negative values of $g$ to an effective weakly interacting model. 
Here, $x_{i\leq N_\downarrow}$ ($x_{i> N_\downarrow}$) are the coordinates of the spin-down (spin-up) fermions; $t_{x_i}=-\hbar^2 \partial_{x_i}^2/(2m)$ is the kinetic energy operator and $v_{\text{ext}}$ describes an external trap. 
The Hamiltonian in \cref{eq:GY Hamiltonian} with a harmonic confinement is of particular interest in cold-atom physics where it can be realized experimentally~\cite{4GaudinYang,Sowinski_2019,MISTAKIDIS20231}. 
Its strongly repulsive limit is well understood by now~\cite{Minguzzi2022,MISTAKIDIS20231}. Here, we discuss the limit of strong attractions, which received less attention. 

Without loss of generality, we assume that $N_\uparrow \geq N_\downarrow $.
In the strongly attractive regime opposite-spin fermions form deeply bound dimers whose binding energy $B_\text{D}\simeq\hbar^2\tilde{g}^2/(4m)$ (see, e.g., Ref.~\cite{4GaudinYang}) is much larger than any other energy scale of the problem~\footnote{A large binding energy implies that the dimer cannot be broken in scattering processes and can be treated as a single degree of freedom. For systems with a finite density, $n$, this binding energy should be compared with typical kinetic energies that are given by $\hbar^2 n^2/m$. Our effective theory is applicable when $n/|\tilde g|\ll1$. As will become clear in the following, the difference between exact energies and effective theory results is less than one percent when $n/|\tilde g|<0.04$.}, here $\tilde{g}:=m g/\hbar^2$.
At the same time, the size of the dimer, $r_\text{D}=-1/\tilde g$~\footnote{We define $r_\text{D}$ as the length scale that contains most of the probability to find two fermions, forming a bound dimer, close to each other, i.e., $\int_{-r_\text{D}}^{r_\text{D}}\Phi_\text{D}(z)^2\text{d}z= 1-e^{-1}\approx0.63$, where $\Phi_\text{D}\propto\exp{\{-|\tilde{g}z|/2\}}$ is the wave function of the dimer.}, is much smaller than any other length scale~\footnote{More precisely, a strongly attractive regime in our work implies that the size of the dimer and hence the effective range is unresolved in scattering, i.e., $k r_\text{D}\ll1$ for all relevant values of the momentum $k$. }. 
This provides a separation of scales for constructing an effective theory with $N_{\downarrow}$ dimers and $M=N_\uparrow-N_{\downarrow}$ unpaired fermions~\footnote{Other possibilities appear impossible because
identical fermions cannot form bound states with more than two particles. This follows from few-body calculations~\cite{Kartavtsev2009,23tononi2022binding} or distribution of momenta in the GY model~\cite{Takahashi1971,6takahashi2005thermodynamics}. Although these considerations are based on a homogeneous geometry, the conclusion holds also for the external potentials $v_{\text{ext}}$ that change slowly on the length scales given by the size of the dimer state.}:
\begin{equation}
h=\sum_{i=1}^{N_\downarrow}\left[\frac{t_{y_i}}{2}+2v_\text{ext}(y_i)\right]+\sum_{i=1}^{M}\left[t_{z_i}+v_\text{ext}(z_i)\right]+w,
\label{eq:ham_effective}
\end{equation}
where the first term describes the dimers that have mass $2m$ and feel a stronger external potential due to their composite nature; 
$w$ is the effective interaction derived below by considering \cref{eq:GY Hamiltonian} for three and four fermions on a line (i.e., with $v_\text{ext}=0$). The low energy spectrum of the Hamiltonian $H$ plus the dimer binding energies $N_{\downarrow}B_\text{D}$ gives the spectrum of $h$.

{\it Three and four fermions on a line.---} In the absence of confinement ($v_{\text{ext}}=0$), the problem is solvable using the Bethe ansatz~\cite{10yang1967some,11gaudin1967systeme,47bethe1931theorie,supp}
allowing us to extract the scattering information for the effective degrees of freedom (dimers and fermions). 
For distinguishable particles, scattering properties in 1D follow from the wave function
\begin{align}
     \Psi(x_\text{rel})= e^{ik x_\text{rel}}+\left(f_\text{e}+f_\text{o}\frac{|x_\text{rel}|}{x_\text{rel}}\right)e^{ik |x_\text{rel}|}\ ,
     \label{eq:scattering_1D}
\end{align}
where $|x_\text{rel}|\to\infty$ describes the relative distance between the particles~\footnote{In the situation where the dimer consists of fermions 1 and 2 the relative distance reads $x_\text{rel}=(x_1+x_2)/2-x_3$. In the case of two dimers made of fermions 1 \& 3 and 2 \& 4, respectively, the relative distance reads $x_\text{rel}=(x_1+x_3)/2-(x_2+x_4)/2\ $.}; $k$ is the corresponding momentum; $f_\text{e}$ and $f_\text{o}$ are, respectively, the even- and odd-channel scattering amplitudes. They enjoy the following effective range expansions as $k\to0$~\cite{18hammer2010causality}
\begin{align}
    \frac{1}{f_\text{e}}\simeq -1-ik a_\text{e}-i\dfrac{r_\text{e} k^3}{2}, \;\;\; 
    \frac{1}{f_\text{o}}\simeq -1-i\dfrac{a_\text{o}}{ k}-i\dfrac{r_\text{o} k}{2}\label{eq: ERE}\ ,
\end{align}
with the scattering length $a_\text{e/o}$ and the effective range $r_\text{e/o}$. By comparing the Bethe ansatz solution with Eqs.~(\ref{eq:scattering_1D}) and~(\ref{eq: ERE}), we extract parameters for fermion-dimer (FD) and dimer-dimer (DD) scattering~\cite{supp}:
\begin{align}
    \text{FD: }&a^\text{FD}_\text{e}=r^\text{FD}_\text{o}/2=3r_\text{D}; \qquad r^\text{FD}_\text{e}=0; \;\; a^\text{FD}_\text{o}=0.\label{eq:FD_scattering}\\
    \text{DD: }&a_\text{e}^\text{DD}=r_\text{D}; \qquad \qquad \qquad r_\text{e}^\text{DD}=0; \;\;  f_\text{o}=0. \label{eq:DD_scattering}
\end{align}
As the dimers obey bosonic statistics, their scattering occurs solely in the even channel. By contrast, the fermion-dimer system can interact in both channels. In the strongly interacting limit $r_\text{D} \to 0$, the dimers fully reflect and become impenetrable. The fermion-dimer scattering becomes fully transparent, yet with a $\pi$ phase shift (the signs of the transmitted and incoming waves are different).
The parameters for dimer-dimer scattering are known~\cite{14bmora2005four} and from now on we mainly focus on the fermion-dimer scattering to establish $w$ for Eq.~(\ref{eq:ham_effective}).

In a homogeneous setting, scattering in odd and even channels can be treated independently because parity is conserved. This is also the case for external potentials that change weakly on the length scale given by $r_\text{D}$. In light of this, we introduce parity-conserving effective interactions that reproduce the calculated effective-range parameters. For the even channels, we employ the $\delta$-function
\begin{align}
\begin{split}
V^\text{FD}_\text{e}&=\frac{g}{2}\delta(x_\text{rel})\Longleftrightarrow \partial_{x_\text{rel}}\Psi\bigl|^{x_\text{rel}=0^+}_{x_\text{rel}=0^-}=\dfrac{g\mu_\text{FD}}{\hbar^2}\Psi(0),\\
V^\text{DD}_\text{e}&=g\delta(x_\text{rel})\Longleftrightarrow \partial_{x_\text{rel}}\Psi\bigl|^{x_\text{rel}=0^+}_{x_\text{rel}=0^-}=\dfrac{2g\mu_\text{DD}}{\hbar^2}\Psi(0)\ ,
\end{split}\label{eq:even channel eft+bc}
\end{align}
where the right-hand-sides represent the interactions as boundary conditions; $\mu_\text{FD}=2m/3$ and $\mu_\text{DD}=m/2$ are the relevant reduced masses.
By inserting a piecewise-defined plane wave for $x_\text{rel}<0$ and $x_\text{rel}>0$, one can show that the effective interactions lead to  \cref{eq:FD_scattering}. One remark is in order here, $g$ is negative in our system, and naively Eq.~\ref{eq:even channel eft+bc} seem to suggest presence of bound states in dimer-dimer and fermion-dimer systems. In reality, these bound states are not present and should be eliminated as we exemplify below. This omission of bound states is natural in our effective model because these states contain momenta that are beyond our low-energy description, see also Ref.~\cite{14bmora2005four} where dimer-dimer bound states are discussed.

Compared to the rather simple form of Eq.~\ref{eq:even channel eft+bc}, the odd-channel interaction of the fermion-dimer system has a more complicated structure. Indeed, scattering in this case is dominated by the non-vanishing effective range $r_\text{o}^\text{FD}$~\footnote{The vanishing scattering length $a_\text{o}^{\text{FD}}$ implies the existence of a zero-energy virtual state~\cite{22barlette2000quantum}, which in the fermion-dimer system corresponds
to a trimer state of negative parity at the threshold for binding~\cite{Kartavtsev2009,23tononi2022binding}.}, beyond the standard odd-channel zero-range potentials~\cite{56cheon1999fermion,55girardeau2004theory}.
Arguably, the simplest finite-range potential~\footnote{The use of finite-range potentials is in general necessary to satisfy the Wigner lower limit~\cite{Wigner1955} 
for one-dimensional scattering phase shifts. In detail the Wigner bound implies that there can be no causality-preserving zero-range interaction for a non-vanishing effective range $r_\text{o}\neq0$~\cite{18hammer2010causality}} to model the odd-channel FD scattering is an attractive square well whose strength is $-V_0$ if $|x_\text{rel}|<r^{\text{FD}}_\text{o}$  and zero otherwise.
The parameter $V_0=\pi^2 \hbar^2/(8 \mu_{\text{FD}} (r^{\text{FD}}_\text{o})^2)$ is tuned such that $a^\text{FD}_\text{o}=0$~\cite{supp}. 

Note that the size of the well vanishes in the limit  $\tilde{g}\to-\infty$ ($r^\text{FD}_\text{o}\simeq -6/\tilde{g}$, see Eq.~\ref{eq:FD_scattering}); it cannot be resolved for low energies, $kr^\text{FD}_\text{o} \to 0$. This allows us to reframe the problem in terms of a boundary condition that fixes the wave function outside the well:
\begin{align}
    V^\text{FD}_\text{o}&\Longleftrightarrow \partial_{x_\text{rel}}^{2}\Psi\bigl|^{x_\text{rel}=0^+}_{x_\text{rel}=0^-}=\dfrac{g \mu_\text{FD}}{\hbar^2}\partial_{x_\text{rel}}\Psi(x_\text{rel}=0)\ .\label{eq:odd channel eft+bc}
\end{align}
Although this potential has vanishing range by construction, which is at odds with the Wigner lower limit~\footnote{As this boundary condition corresponds to a zero-range interaction, it must violate causality for $1/\tilde g\neq 0$ as the Wigner lower limit indicates for $r_\text{o}\neq 0$. This manifests itself as non-hermiticity of the resulting problem, which can be demonstrated by non-orthogonality of the odd solutions of a two-body problem. As we illustrate in the Letter, this does not preclude us from estimating the energies in the first order of $1/\tilde g$~\cite{supp}.}, it leads to the correct energies of the fermion-dimer system in the limit $1/g\to0$ as we illustrate below~\cite{supp}.  Additionally, this boundary condition will allow us to map the effective theory onto a weakly interacting system~\footnote{We remark here that the boundary conditions introduced in \cref{eq:odd channel eft+bc} do not correspond to the odd channel interaction $\delta^\prime$ introduced in Ref.~\cite{56cheon1999fermion}. Indeed, the potential $\delta^\prime$ connects the derivative of the wave function to the discontinuity of the wave function itself, whereas \cref{eq:odd channel eft+bc} connects the derivative of the wave function to the discontinuity of the second derivative of the wave function. It is worth noting however that in the limit $1/g=0$, both \cref{eq:odd channel eft+bc} and $\delta^\prime$ demand that the derivative of the wave function vanishes, which leads to the important conclusion that strongly-interacting odd channel can be mapped onto a weakly-interacting even channel~\cite{57granger2004tuning}. }. To understand the limits of validity of our effective theory formulation, we consider a three-body problem below. An analysis of a four-body problem yields similar conclusions~\cite{supp}. 

\begin{figure}
    \centering
    \includegraphics[width=0.48
    \textwidth]{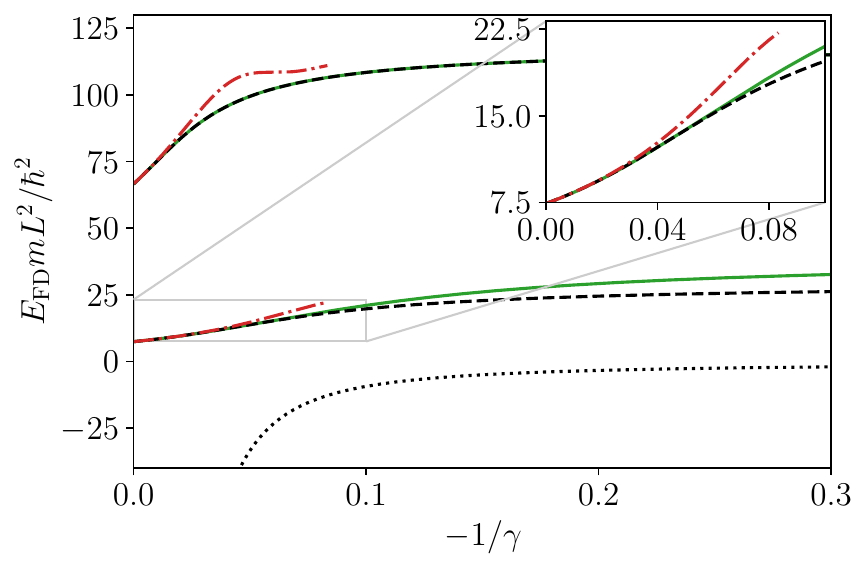}
    \caption{Energy of the effective fermion-dimer system on a ring. Green solid curves show the corresponding energies of the GY model. Red dash-dotted curves are for a square well potential. For $\gamma=-12$, the size of this potential approaches the length of the ring causing strong finite-size effects. Black dashed curves show the energy within the effective theory (solution of \cref{eq:EFT ring momenta eq}). The lowest (dotted) curve is an artifact of the effective model and should be omitted. Inset presents the region of strong interactions for the ground state.}
    \label{fig: FDEFT vs GY model}
\end{figure}

{\it Three fermions on a ring. ---}
First, we consider three fermions on a ring of length $L$. This system is one of the simplest instances of the Gaudin-Yang (GY) model~\cite{11gaudin1967systeme,10yang1967some}, providing us with the exact results for benchmarking the performance of $V^\text{FD}_\text{e}$ and $V^\text{FD}_\text{o}$. For simplicity, we work with vanishing total momentum, $P=0$, here~\cite{supp}.

Using the boundary conditions~(\ref{eq:even channel eft+bc}) and~\ref{eq:odd channel eft+bc}), we find the equation for the energies of the fermion-dimer system~\cite{supp}
\begin{equation}
    3kL\tan(kL/2)=\gamma ,\label{eq:EFT ring momenta eq}
\end{equation}
where $\gamma=\tilde{g}L$ is the dimensionless interaction strength and $k$ is the relative momentum. We plot the effective fermion-dimer energy $E_\text{FD}=\hbar^2k^2/2\mu_\text{FD}$ in \cref{fig: FDEFT vs GY model} together with the exact three-body energy calculated using the GY model, according to the prescription $E_\text{FD}=E^{\text{GY}}_\text{3-body}+B_\text{D}$. Note that \cref{eq:EFT ring momenta eq}
is applicable for both even and odd channels.
The underlying physical reason for the degeneracy of the energy levels is the equivalence of the left and right directions in a ring geometry. We obtain another solution if we change the sign of momenta in the Bethe ansatz wave function~\cite{supp}.

Disregarding the deep bound state, which does not exist in the GY model and should be omitted, Fig.~\ref{fig: FDEFT vs GY model} demonstrates that \cref{eq:EFT ring momenta eq} describes the exact energies for $-1/\gamma\lesssim 0.04$ well (the largest relative deviation in this region is less than one percent~\footnote{This numerical estimate is in agreement with the construction of our model, exact in the order $1/\gamma$. A detailed investigation of the beyond $1/\gamma$-physics is outside the scope of the present Letter. We remark however that in some cases the effective model is accurate even at the level $1/\gamma^2$. One example is a spin-balanced limit of the GY model. Indeed, the approximate energy of the Lieb-Liniger gas of dimers~\cite{LLTDresultMinguzzi} $
\frac{E_0}{NE_\text{F}}\simeq \frac{1}{12}\left[1-\frac{\hbar^2 N}{m g L}+\frac{3 (\hbar^2 N)^2}{4(m g L)^2}\right]$ is in agreement with the direct solution of the GY model~\cite{21fuchs2004exactly}. Another example is actually the present fermion-dimer system, see S.4. of \cite{supp}.}).
 In \cref{fig: FDEFT vs GY model}, we also present the energies for an attractive square well in place of the fermion-dimer potential. For large values of $|\gamma|$, the corresponding energy spectrum is described by \cref{eq:EFT ring momenta eq}. For $-1/\gamma\simeq 0.1$, the size of the square well becomes comparable to the length of the ring and the finite-size effects cannot be neglected~\cite{supp}.

{\it Three fermions in a harmonic trap.---} The fact that the boundary conditions~(\ref{eq:odd channel eft+bc}) work well in a ring might appear natural because we used a homogeneous system to build them. Therefore, as a next illustration, we work with three particles in a harmonic trap, i.e., with $v_\text{ext}(x)=m\omega^2x^2/2$. This potential sets the 
length scale $l=\sqrt{\hbar/(m\omega)}$ suggesting the following definition for a dimensionless interaction strength: $\gamma^\text{HO}=g/(\hbar\omega l)$. 

\begin{figure}
    \centering
    \includegraphics[width=0.48
    \textwidth]{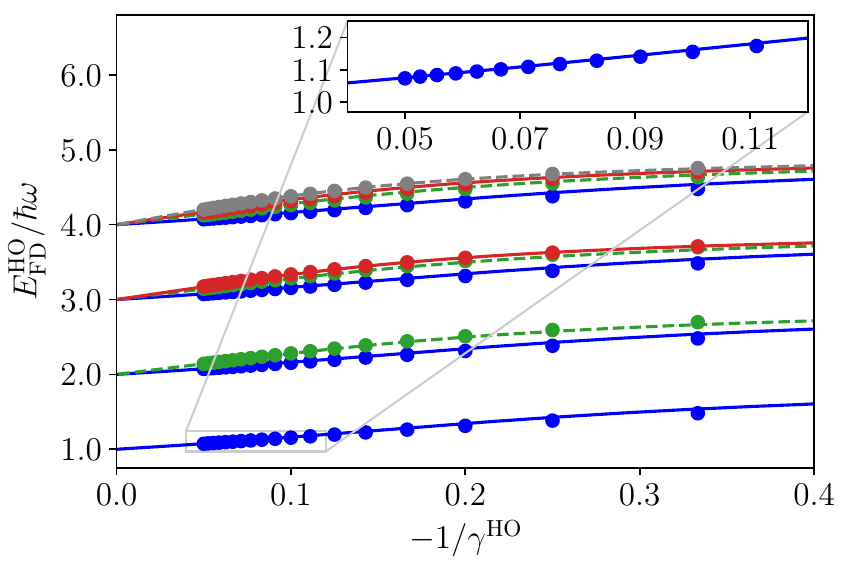}
    \caption{Energy of the effective fermion-dimer system in a harmonic trap. Results of \cref{eq:HOEFTbc} are shown for odd (solid curves) and even states (dashed curves). Markers present outcomes of the numerical transcorrelated method. Curves with identical colors represent states with the same relative motion. Inset presents the region of strong interactions for the ground state.}
    \label{fig: HOFDTCMenergie}
\end{figure}

The solution of the fermion-dimer model in a harmonic oscillator potential is straightforward after one notices that the center-of-mass motion can be decoupled from the relative one (a general feature of harmonically interacting systems~\cite{Armstrong2011}).
The energy of the system is 
\begin{align}
E^\text{HO}_\text{FD}=\hbar\omega\left(2\nu_j+n_\text{COM}+1\right), \label{eq:EHOFD}
\end{align}
where $n_\text{COM}$ is an integer that determines the center-of-mass dynamics.  The quantum numbers $\nu_j$ for the relative motion are found from the equations~\cite{supp}:
\begin{align}
     \dfrac{l_\text{FD}}{l}=-\dfrac{\gamma^\text{HO}\Gamma\left(-\nu_\text{e}\right)}{4\Gamma\left(-\tilde \nu_\text{e}\right)},\;
    \;\; \dfrac{l_\text{FD}}{l}=\dfrac{\gamma^\text{HO}\Gamma\left(-\tilde \nu_\text{o}\right)}{\left(4\nu_\text{o}+1\right)\Gamma\left(-\nu_\text{o}\right)}\ ,\label{eq:HOEFTbc}
\end{align}
where $\tilde \nu_j=\nu_j-1/2$ and 
$l_\text{FD}=\sqrt{\hbar/(\mu_\text{FD}\omega)}$.
The trap breaks translational invariance of the problem and thus lifts the degeneracy between even and odd solutions~\footnote{Contrast also the uniqueness of the ground state for $1/\gamma^{\text{HO}}\to 0^-$ 
with triple degeneracy of the ground state in the limit $1/\gamma^{\text{HO}}\to 0^+$ ~\cite{Guan2009,Gharashi2013,Volosniev2013}.
This illustrates the fact that one cannot consider orderings of particles (e.g, $x_1<x_2<x_3$) as independent for strongly attractive systems.}, which was present in the GY model, see \cref{eq:EFT ring momenta eq}.

To benchmark \cref{eq:HOEFTbc}, we diagonalize the Hamiltonian in \cref{eq:GY Hamiltonian} for a three-body system numerically. In spite of a small number of particles, diagonalization of $H$ is virtually impossible using standard methods already for intermediate interaction strengths (e.g., $\gamma^{\text{HO}}\simeq  -5$)~\cite{61d2014three,1rammelmuller2023magnetic}. Therefore, we resort to a transcorrelated method where the leading-order singularity due to the boundary condition is removed via a similarity transformation~\cite{68jeszenszki2018accelerating,Jeszenszki2020}. The computations were performed using the open-source package \texttt{Rimu.jl} \cite{rimucode,supp}.
In \cref{fig: HOFDTCMenergie}, we compare the energy $E^{\text{TCM}}_\text{3-body}+B_\text{D}$ with the prediction of \cref{eq:EHOFD}. The overall agreement for the considered values of $\gamma^{\text{HO}}$ for the ground as well as low-lying excited states provides a further validation of the proposed effective theory. Note that the energy spectrum for $1/\gamma^{\mathrm{HO}}=0$ coincides with that of two non-interacting particles, the physical reason for that is explained below. 

{\it Many-body problem in a trap. ---} The discussion above establishes the Hamiltonian from Eq.~(\ref{eq:ham_effective}) as a viable framework for studying strongly attractive 1D fermions. Further progress can be made by mapping Eq.~(\ref{eq:ham_effective}) onto a weakly-interacting model. To implement this mapping, we rely on the boundary conditions from \cref{eq:even channel eft+bc,eq:odd channel eft+bc}. Let us assume that we have access to an eigenfunction $\Psi$ of $h$ for a given external potential. This function is defined on orderings of effective degrees of freedom, e.g., $y_1<y_2<...<z_M$. 
Now, for a fixed ordering of fermions $z_1<z_2<...<z_M$, we construct a new function $\phi=(-1)^{\text{sign}(P)}\Psi$, where $\text{sign}(P)$ is the parity of the permutation $P$ in a set of coordinates of dimers and fermions. For example, $\phi=\Psi$ for the ordering $y_1<y_2<...<z_M$, whereas 
$\phi=-\Psi$ for the ordering where the first two dimers are exchanged, i.e., $y_2<y_1<...<z_M$. The function $\phi$ now can be trivially extended to any ordering of $\{z_1,...,z_M\}$ using fermionic symmetry \cite{58girardeau2007soluble}.

Since we use boundary conditions to define the effective interactions $w$, the function $\phi$ is (by construction) an eigenstate of the operator $h-w$ from \cref{eq:ham_effective} for every ordering of particles just as $\Psi$. However, boundary conditions for $\phi$ differ from Eqs.~(\ref{eq:even channel eft+bc}) and (\ref{eq:odd channel eft+bc}).   
For the odd channel, they are constructed following Refs.~\cite{56cheon1999fermion,55girardeau2004theory} from the rules for even-channel scattering in the original picture. For the dimer-dimer and fermion-dimer scattering we have 
\begin{align}
\begin{split}
\phi\bigl|^{x_\text{rel}=0^+}_{x_\text{rel}=0^-}= \frac{{2\hbar^2}}{\mu_\text{DD} g}
\partial_{x_\text{rel}}\phi (x_\mathrm{rel}=0), \\ \phi\bigl|^{x_\text{rel}=0^+}_{x_\text{rel}=0^-}= \frac{{4\hbar^2}}{\mu_\text{FD} g}
\partial_{x_\text{rel}}\phi(x_\mathrm{rel}=0).
\end{split}
\end{align}
For the even-channel fermion-dimer scattering, instead of \cref{eq:odd channel eft+bc} the following condition must be satisfied 
\begin{align}
\partial_{x_\text{rel}}\phi\bigl|^{x_\text{rel}=0^+}_{x_\text{rel}=0^-}=\frac{4\hbar^2}{\mu_{\text{FD}}g}\partial_{x_\text{rel}}^{2}\phi(x_{\mathrm{rel}}=0).
\end{align}
In the limit $g\to-\infty$, the wave function features no peculiarities, and corresponds to a non-interacting mixture of two mass-imbalanced Fermi gases~\footnote{This result can be benchmarked against a Bethe ansatz solution for a system in a box trap of length $a$.  Indeed, the energy of the non-interacting mixture in a box trap is given by $\frac{\hbar^2}{24m}\frac{\pi^2}{a^2}M(M+1)(2M+1)+\frac{\hbar^2}{12m}\frac{\pi^2}{a^2}N_{\downarrow}(N_{\downarrow}+1)(2N_{\downarrow}+1)$ in agreement with Ref.~\cite{5oelkers2006bethe}.}. 

\begin{figure}
    \centering
    \begin{minipage}{0.48\textwidth}
        \centering
        \includegraphics[width=\textwidth]{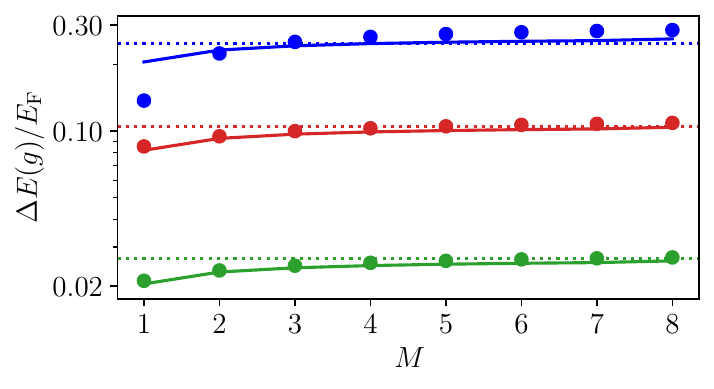}
    \end{minipage}
    \caption{Energy of the dimer confined within a harmonic trap, $\Delta E(g)=E(g)-E(g\to-\infty)$, as a function of the number of unpaired fermions $M=N_{\uparrow}-1$ for different interaction strengths $\gamma_\text{scaled}\in\{-10,-25, -100\}$ (blue, red, green). Symbols show numerical results while solid lines correspond to a perturbative calculation, both within effective theory, see the text for details. Note that we use the Fermi energy $E_\text{F}=(M+1/2)\hbar\omega$ as a unit of energy here; $\gamma_\text{scaled}=\gamma^\text{HO}\pi/\sqrt{2M}$.
      Dotted lines are the many-body energies calculated using the Bethe ansatz, see~\cite{supp} for details.}
    \label{fig: ThermoLimit}
\end{figure}

One important advantage of working with a weakly-interacting system is the opportunity to use perturbation theory. The ground state of $h$ is non-degenerate, its energy in the $1/g$-th order reads 
\begin{equation}
E_N=E^\text{F}_{M}+E^\text{F}_{N_{\downarrow}}+\langle \phi| V_{\text{FD}}+ V_{\text{DD}}|\phi\rangle,
\label{eq:energy_many_body}
\end{equation}
where the first two terms on the right hand side are the non-interacting energies for $M$ unpaired fermions and $N_{\downarrow}$ dimers. The last term represents the first-order correction due to the interaction. It can be easily calculated as it represents a sum of two-body corrections \cite{supp}. 

To illustrate this, we consider a problem of a single impurity in a Fermi gas (i.e., $N_\downarrow=1$) confined in a harmonic trap.
 For repulsive interactions, this problem was extensively studied numerically~\cite{Astrakharchik2013,Lindgren2014,Gharashi2015,Pecak2016} and realized experimentally in few-atom systems~\cite{Wenz2013}. In the limit $g\to-\infty$, this problem transforms into another impurity problem: a dimer interacting with $M=N_{\uparrow}-1$ fermions. The corresponding ground state energy is given by \cref{eq:energy_many_body} with $\langle \phi| V_{\text{DD}}|\phi\rangle=0$ and
$\langle \phi| V_{\text{FD}}|\phi\rangle=\sum_{n=0}^{M} \Delta E_2(n)$, where $\Delta E_2(n)$ is the energy shift due the interaction in a two-body problem assuming that the non-interacting system has a fermion in the $n$th one-body energy level of the trap~\cite{supp}. Figure~\ref{fig: ThermoLimit} shows this perturbative result~\cite{supp}. 
We rescale the energy in the units of the Fermi energy $E_\text{F}$ for a faithful comparison of systems with different values of $M$. 
The dimensionless interaction strength is given by $\gamma_\text{scaled}=\gamma^\text{HO}\pi/\sqrt{2M}$.

 The energies in Fig.~\ref{fig: ThermoLimit} quickly approach the thermodynamic limit (i.e., $M\to \infty$ with fixed $E_\text{F}$), where the problem is often called the Fermi polaron~\cite{MISTAKIDIS20231}.  
Our framework allows one to calculate this approach, and together with the known results for repulsive interactions ~\cite{Wenz2013,Levinsen2015} provides an intuitive picture of a few-to-many-body crossover in strongly interacting 1D systems with impurities. Note that we can study this crossover with a handful fermions. Indeed, for $M\gtrsim5$ we calculate $\lim_{|\gamma_{\mathrm{scaled}}|\to\infty}|\gamma_\text{scaled}| \Delta E_{M}/E_\text{F}\simeq 2.6$. This equation determines the energy as well as the universal tail
of the momentum distribution (contact parameter)~\cite{Tan2008,Barth2011}.
For $M\to\infty$ the role of the trap is negligible and we derive from the Bethe ansatz~\cite{8mcguire1966interacting,supp}: $\lim_{|\gamma_{\mathrm{scaled}}|\to\infty}|\gamma_\text{scaled}| \Delta E_{M}/E_\text{F}=8/3$, in agreement with our few-body result.

To illustrate the finite-range effects on the energies, Fig.~\ref{fig: ThermoLimit} also presents results of a direct numerical diagonalization of the effective model in Eq.~(\ref{eq:ham_effective}) with Eq.~(\ref{eq:even channel eft+bc}) and a square-well potential in place of an odd-channel fermion-dimer interaction~\cite{supp}. 
This energy agrees with the perturbative result of Eq.~(\ref{eq:energy_many_body}) for $\gamma_{\text{scaled}}=-25$ and $\gamma_{\text{scaled}}=-100$.  The deviation between the two methods at $\gamma_{\text{scaled}}=-10$ signals a departure from the leading order in~$1/g$. We observed that this deviation becomes less pronounced as the number of particles increases. Our interpretation is that by fixing $E_\text{F}$, we effectively increase the size of the trap when increasing $M$. This reduces the finite-size effects, which dominate physics for the corresponding values of $g$, see Fig.~\ref{fig: FDEFT vs GY model}.

{\it Summary.---} We studied a two-component fermionic system with strong attractive particle-particle  interactions. We showed that this model can be mapped onto a weakly-interacting  mass-imbalanced fermionic mixture, which can be analyzed using many-body perturbation theory. Our results rely on the separation of scales, $B_\text{D}/(\text{Kinetic}\; \text{Energy})\gg 1$, making them applicable to dilute samples at zero or low temperatures. This condition is also satisfied for low-energy collective excitations, so our effective model can be directly used to study real-time dynamics of strongly attractive particles.
The proposed framework could be extended to other systems, in particular, $SU(N)$ fermionic mixtures that can be realized in a cold-atom laboratory~\cite{Pagano2014}. These systems are Bethe ansatz solvable in the homogeneous case. Hence, we conjecture that in the limit of strong attractive interactions, such systems can be mapped onto weakly-interacting mass-imbalanced models whose constituents are the available bound states. For $SU(3)$, e.g., the constituents would be trimers, dimers, and fermions~\footnote{The conjecture follows from contradiction: Let us assume that it is not possible to map such a system onto a weakly-interacting mass-imbalanced model. This implies the existence of a Bethe-ansatz solvable mass-imbalanced system, since the underlying GY model is Bethe-ansatz solvable with the wave function describing an effective system of constituents (unpaired fermions, dimers, trimers, etc.) in the limit of strong attractive interactions. However, mass-imbalanced systems are not Bethe-ansatz solvable, see Refs.~\cite{Lamacraft2013,Huber2021} illustrating this point using the smallest three-body problem.}.

\acknowledgements
We thank Nathan Harshman and Fabian Essler for useful discussions. Further, we thank Stephanie Reimann and the Lund cold atom group for giving us access to their configuration interaction code~\cite{CremonPhDThesis, BjerlinPhdThesis}. This work has been
supported by the Deutsche Forschungsgemeinschaft (DFG, German Research Foundation) – Project-ID 279384907 – SFB 1245 (H.W.H. and T.G.B.);  by the Marsden Fund of New Zealand (Contract No.\ MAU 2007), from government funding managed by the Royal Society of New Zealand Te Ap\=arangi (J.B.\ and M.Č.); by the New Zealand - eScience Infrastructure (NeSI) high-performance computing facilities in the form of a merit project allocation;
by the Danish National Research Foundation through the Center of Excellence ``CCQ'' (DNRF152) (A.G.V.).

\section*{data availability}
The data supporting the findings of this Letter are available in a public Zenodo repository at \cite{paperdata}.\\ \\ \\ \\ \\ \\ \\ \\ \\ \\ \phantom{.}

\widetext

\renewcommand{\theequation}{S\arabic{equation}}
\renewcommand{\thefigure}{S\arabic{figure}}
\renewcommand{\thesection}{S.\arabic{section}}
\section*{\texorpdfstring{Supplemental Material for\\ ``Effective theory for strongly attractive one-dimensional fermions''}{Supplemental Material}}
The supplemental material provides technical details behind the findings of the main text. 
Certain derivations, e.g., the Bethe-ansatz equations can be found elsewhere. We present them mainly for convenience of the reader.
The supplemental material is organized as follows: In section~\ref{sec:S1BA}, we discuss the Bethe ansatz solution for three and four fermions. In section~\ref{S2pbcs}, we consider  a mass-imbalanced system in a ring. In section~\ref{S3EFTHO}, we solve dimer-dimer and fermion-dimer problems in a harmonic trap. In sections~\ref{appendix:fsw} and ~\ref{S5ESWHO}, we compare and contrast the effective interaction in the form of a boundary condition with a square well potential.
In section~\ref{S6Manybody}, we review the exact-diagonalization approach, which was used to calculate the energies of a spin-polarized Fermi gas with an impurity. In section~\ref{S7TCM}, we outline the transcorrelated method, which was used to benchmark our effective theory in harmonic confinement for a fermion-dimer system. In section~\ref{S8Manybodyperturbation}, we discuss perturbation theory for a many-body problem. \\ \\
We refer to equations from the main part as Eq.~($n$) and to equations of the supplement as Eq.~(S$n$); similar holds for the figures. The data supporting the findings of this Supplemental Material are available in a public Zenodo repository at \cite{paperdata}.

\vspace*{1em}

\section{Bethe ansatz solution}
\label{sec:S1BA}

To find the scattering solution for the system with $N=N_\downarrow+N
_\uparrow$, we use the Bethe ansatz \cite{10yang1967some,11gaudin1967systeme,4GaudinYang,6takahashi2005thermodynamics}
\begin{equation}
\Psi(\{x_i\})=\sum_{P,Q\in S(N)}\Theta(Q,\{x_i\})[P,Q]e^{i\sum_{j}^N k_{P(j)}x_{Q(j)}},\label{eq:BA Appendix}
\end{equation}
where $[P,Q]$ are coefficients, $k_j$ are quasi-momenta, $\Theta$ is an $N$-dimensional Heaviside-function. The sum goes over all elements of the permutation group $S(N)$. Applying the known procedure~\cite{5oelkers2006bethe,6takahashi2005thermodynamics} to our problem, we define the $\Phi$-coefficients  using one-line notation for $Q$
  \begin{align}
     &\begin{aligned}
         &\text{$N_\downarrow=1$; $N_\uparrow=2$:}\\
         &\hspace{1.2cm} \tikzmark{311}\ \Phi(1;P):=\text{sign}(P)[P,(123)]\\
         &\hspace{1.2cm} \tikzmark{312}\ \Phi(2;P):=\text{sign}(P)[P,(312)]\\
         &\hspace{1.2cm} \tikzmark{313}\ \Phi(3;P):=\text{sign}(P)[P,(231)]\\
         &\\
         &\\
         &\\
         \end{aligned}
         &\begin{aligned}
         &\text{$N_\downarrow=2$; $N_\uparrow=2$:}\\ 
         &\hspace{1.2cm} \tikzmark{421}\ \Phi(1,2;P):=\text{sign}(P)[P,(1234)]\\
         &\hspace{1.2cm} \tikzmark{422}\ \Phi(1,3;P):=\text{sign}(P)[P,(1423)]\\
         &\hspace{0.5cm}\hspace{1.2cm} \tikzmark{4221}\ \Phi(2,3;P):=\text{sign}(P)[P,(3124)]\\
         &\hspace{1.2cm} \tikzmark{423}\ \Phi(1,4;P):=\text{sign}(P)[P,(1342)]\\
         &\hspace{0.5cm}\hspace{1.2cm} \tikzmark{4231}\ \Phi(2,4;P):=\text{sign}(P)[P,(4132)]\\
         &\hspace{0.5cm}\hspace{1.2cm}\tikzmark{4232}\ \Phi(3,4;P):=\text{sign}(P)[P,(3412)]\label{eq:def phi}\ ,\\
     \end{aligned}
\begin{tikzpicture}[overlay,remember picture]
		\draw[-{Latex[length=2.5mm]},rounded corners,line width=0.4mm] (311.west)+(-1,0.3)|-(311.north);
		\draw[-{Latex[length=2.5mm]},rounded corners,line width=0.4mm] (311.west)+(-1,0.3)|-(312.north);
		\draw[-{Latex[length=2.5mm]},rounded corners,line width=0.4mm] (311.west)+(-1,0.3)|-(313.north);
		\draw[-{Latex[length=2.5mm]},rounded corners,line width=0.4mm] (421.west)+(-1,0.3)|-(421.north);
		\draw[-{Latex[length=2.5mm]},rounded corners,line width=0.4mm] (421.west)+(-1,0.3)|-(422.north);
		\draw[-{Latex[length=2.5mm]},rounded corners,line width=0.4mm] (421.west)+(-1,0.3)|-(423.north);
		\draw[-{Latex[length=2.5mm]},rounded corners,line width=0.4mm] (422.west)+(0,0.1)|-(4221.north);
		\draw[-{Latex[length=2.5mm]},rounded corners,line width=0.4mm] (423.west)+(0,0.1)|-(4231.north);
		\draw[-{Latex[length=2.5mm]},rounded corners,line width=0.4mm] (423.west)+(0,0.1)|-(4232.north);
	\end{tikzpicture}
\end{align}
where the arguments of $\Phi$ describe the positions of the spin-down fermions within the particle ordering and the considered permutation $P$.  Due to the indistinguishability of same-spin fermions the coefficients $[P,Q]$ change the sign for each exchange of same-spin fermions. This is the reason why in \cref{eq:def phi}, only 3 out of 6 for the $1+2$ system (6 out of 24 permutations for the $2+2$ system) $Q$ are needed for the definition of $\Phi$.
The corresponding Bethe ansatz equations read (with $\tilde{g}=\dfrac{m}{\hbar^2}g$ and spin-roots $\Lambda_j$) \cite{4GaudinYang,6takahashi2005thermodynamics,5oelkers2006bethe}
 \begin{align}
 \begin{split}
    \Phi(\{\tilde{y}_i\};P)=&\sum_{R\in S(N_\downarrow)}A(R)\prod_{i=1}^{N_\downarrow}F(\tilde{y}_i,\Lambda_{R(i)};P)\\
    F(\tilde{y}_i,\Lambda;P):=&\prod_{l=1}^{\tilde{y}_i-1}(k_{P(l)}-\Lambda+i\tilde{g}/2)\prod_{l=\tilde{y}_i+1}^{N}(k_{P(l)}-\Lambda-i\tilde{g}/2)\\
    A(R):=&\text{sign}(R)\prod_{1\leq j< l\leq N_\downarrow}(\Lambda_{R(j)}-\Lambda_{R(l)}-i\tilde{g})\ .\label{eq: BA sol}
     \end{split}
\end{align}
The parameters $\{\tilde{y}_j\}_{1\leq j\leq N_\downarrow}$ represent the position of the spin-down fermions within the particle ordering.

In the following, we consider the solution to fermion-dimer and dimer-dimer scattering on a line. The formation of bound dimers comes from pairing of quasi-momenta, cf. Refs. \cite{4GaudinYang,6takahashi2005thermodynamics,5oelkers2006bethe,14bmora2005four}. For fermion-dimer scattering this pairing reads $k_{1/2}=k_\text{D}/2\mp i\tilde{g}/2$ and $k_3=k_\text{F}$ with dimer and fermion momentum $k_\text{D}$ and $k_\text{F}$. Similarly, for dimer-dimer scattering the pairing reads $k_{1/3}=k_\text{D1}/2\mp i\tilde{g}/2$, $k_{2/4}=k_\text{D2}/2\mp i\tilde{g}/2$. In both cases the spin-roots are related to the dimer momenta via $\Lambda_j=k_{\text{D}j}/2$.
Inserting these quasi-momenta and the spin-roots into \cref{eq: BA sol} gives us the fermion-dimer scattering wave function in relative coordinates (the equation only holds for $|x_\text{rel}/r_\text{D}|\gg1$~\footnote{If the dimer consists of fermions 1 and 2 the relative distance is defined as $x_\text{rel}=(x_1+x_2)/2-x_3$.})
\begin{align}
\begin{split}
\Psi^\text{FD}\propto&\begin{cases}\text{$x_\text{rel}>0$: }(3k_\text{rel}-i\tilde{g})e^{ik_\text{rel}x_\text{rel}}\\
    \text{$x_\text{rel}< 0$: }(3k_\text{rel}+i\tilde{g})e^{ik_\text{rel}x_\text{rel}}\\
    \end{cases}\ \label{eq:FDEFT rel motion wave function}\\
    =& e^{ik_\text{rel}x_\text{rel}}-\dfrac{1}{1-3ik_\text{rel}/\tilde{g}}\left(1+\dfrac{|x_\text{rel}|}{x_\text{rel}}\right)e^{ik_\text{rel}|x_\text{rel}|}
    \end{split}
\end{align}
and the dimer-dimer scattering wave function
(the equation only holds for $|x_\text{rel}/r_\text{D}|\gg1$~\footnote{If the two dimers are made of fermions 1 \& 3 and 2 \& 4, respectively, the relative distance is defined as $x_\text{rel}=(x_1+x_3)/2-(x_2+x_4)/2\ $.})
\begin{align}
\begin{split}
    \Psi^\text{DD}\propto&(k_\text{rel}+i\tilde{g})e^{-ik_\text{rel}|x_\text{rel}|}+(k_\text{rel}-i\tilde{g})e^{ik_\text{rel}|x_\text{rel}|}\\
    \propto&e^{ik_\text{rel}x_\text{rel}}-\dfrac{1}{1-ik_\text{rel}/\tilde{g}}e^{ik_\text{rel}|x_\text{rel}|}+[x_\text{rel}\leftrightarrow-x_\text{rel}]\ .\label{eq:DDEFT rel motion wave function}
\end{split}
\end{align}
Notice that the anti-symmetric fermion exchange characteristic is included in the already factorized dimer bound state wave function.

We note that the GY model solution on a line can be adjusted to the periodic boundary conditions, which lead to the following Bethe-ansatz equations
(see Refs. \cite{4GaudinYang,5oelkers2006bethe,6takahashi2005thermodynamics,10yang1967some})
\begin{align}
\begin{split}
    \forall 1\leq j \leq N:&\ \prod_{\alpha=1}^{N_\downarrow}\dfrac{k_j-\Lambda_\alpha+i\tilde{g}/2}{k_j-\Lambda_\alpha-i\tilde{g}/2}=e^{i k_{j}L}\\
    \forall 1\leq \alpha \leq N_\downarrow:&\ \prod_{j=1}^N\dfrac{k_j-\Lambda_\alpha+i\tilde{g}/2}{k_j-\Lambda_\alpha-i\tilde{g}/2}=-\prod_{\beta=1}^{N_\downarrow}\dfrac{\Lambda_\alpha-\Lambda_\beta-i\tilde{g}}{\Lambda_\alpha-\Lambda_\beta+i\tilde{g}}\ .
    \end{split}\label{eq:GY model BA eq}
\end{align}
These equations fix the quasi-momenta $k_j$ (and spin-roots $\Lambda_l$). The corresponding energy reads
\begin{equation}
    E^\text{GY}=\dfrac{\hbar^2}{2m}\sum_{i=1}^N k_i^2\ .
\end{equation}
Note that the change $k_j\to-k_j$ and $\Lambda\to-\Lambda$ leads to another solution of the Bethe-ansatz equations. This observation explains the degeneracy of the solutions on a ring discussed in the main text.

Finally, let us simplify the Bethe ansatz equations~(\ref{eq:GY model BA eq}) with a fermion-dimer ansatz $k_{1/2}=-k/2\mp i\tilde{g}/2$, $\Lambda=k_\text{D}/2$ and $k_3=k$ (center-of-mass $\sum_j k_j=P=0$). This leads to the equation for the energies of the GY model  in the limit $1/\tilde g\to0$
\begin{equation}
    \dfrac{3kL+i\gamma}{3kL-i\gamma}=e^{i kL}\Longleftrightarrow3kL\tan(kL/2)=\gamma.
\end{equation}
Note that this equation coincides with \cref{eq:EFT ring momenta eq} from the main text, providing another validation for the use of the proposed boundary conditions, see \cref{eq:even channel eft+bc,eq:odd channel eft+bc} of the main text.

\section{Fermion-dimer system in a ring}\label{S2pbcs}

First of all, we show that the transformation to the center-of-mass frame in a ring is useful for a mass-imbalanced system (like the fermion-dimer system) only for certain values of the total
 momentum $\sum_{j}k_jL=:P=2\pi n$ for $n\in\mathbb{Z}$. To this end, we consider the transformation
\begin{equation}
    \Psi(y,z)=e^{i P \dfrac{m_\text{D}y+mz}{m_\text{D}+m}}\Phi(y-z).
\end{equation}
Periodic boundary conditions $\Psi(y+L,z)=\Psi(y,z)$ and $\Psi(y,z+L)=\Psi(y,z)=\Psi(y,z-L)$ set the following conditions for the function $\Phi$ depending on relative coordinate $x_\text{rel}=y-z$
\begin{align}
    \Phi(x_\text{rel})&=e^{i P \dfrac{m}{m+m_\text{D}}L}\Phi(x_\text{rel}+L)\label{eq:pbc Com1}\\
    \Phi(x_\text{rel})&=e^{-i P \dfrac{m_\text{D}}{m+m_\text{D}}L}\Phi(x_\text{rel}+L)\label{eq:pbc Com2}\ .
\end{align}
These equations are consistent if
\begin{align}
    1=e^{i P \cancelto{1}{\dfrac{m+m_\text{D}}{m+m_\text{D}}}L\hspace{1mm}}\Longleftrightarrow P L=2\pi n \text{ with }n\in\mathbb{Z}\ ,
\end{align}
which correspond to the free motion of the center-of-mass with momentum $P$. Equations~(\ref{eq:pbc Com1}) and~(\ref{eq:pbc Com2}) lead to the common periodic boundary conditions $\Phi(x_\text{rel})=\Phi(x_\text{rel}+L)$ only if a) $P=0$, b)  certain values of $P$ if $\frac{m}{m+m_\text{D}}$ and $\frac{m_\text{D}}{m+m_\text{D}} \in \mathbb{Q}$. Further, the boundary conditions in \cref{eq:pbc Com1,eq:pbc Com2} decouple even and odd channel only if $P \frac{m}{m+m_\text{D}}L,P \frac{m_\text{D}}{m+m_\text{D}}L\in 2\pi\mathbb{Z}$, in which case the boundary conditions reduce to the common one. Based on this insight and since our effective interactions are constructed in even and odd channels, we use our effective interaction formulations for the FD system ($m_\text{D}=2m_\text{F}$) in periodic confinement for $P=0$ (in general, $P=6\pi n$ for $n\in\mathbb{Z}$ can be used).

For $P=0$, two-body problems are decoupled in even and odd channels. We solve them independently using the interaction boundary conditions, see \cref{eq:even channel eft+bc,eq:odd channel eft+bc} of the main text. The even and odd wave functions, which for vanishing total momentum $P=0$ fulfill the periodic boundary condition $\Phi(x_\text{rel}-L/2)=\Phi(x_\text{rel}+L/2)$, read
\begin{align}
\begin{split}
     \Psi_\text{e/o}\propto&\begin{cases}\text{$x_\text{rel}>0$: }e^{ik_\text{rel}x_\text{rel}}+Be^{-ik_\text{rel}x_\text{rel}}\\
    \text{$x_\text{rel}<0$: }\pm\left[Be^{ik_\text{rel}x_\text{rel}}+e^{-ik_\text{rel}x_\text{rel}}\right]
    \end{cases}\text{ with } B=e^{ik_\text{rel}L}\ .
    \end{split}\label{eq:odd sol ring}
 \end{align}
The corresponding momentum is obtained from 
 \begin{equation}
    3ik_\text{rel}L\dfrac{1-e^{ik_\text{rel}L}}{1+e^{ik_\text{rel}L}}=3k_\text{rel}L\tan(k_\text{rel}L/2)=\gamma\ \label{eq:deltaprime ring energy fix}
\end{equation}
which is \cref{eq:EFT ring momenta eq} of the main text.

\section{Two effective degrees of freedom in a harmonic oscillator}\label{S3EFTHO}

{\it Fermion-dimer system.}
The effective two-body Hamiltonian in a harmonic trap allows for decoupling of center-of-mass and relative motions 
\begin{align}
\begin{split}
H^\text{GY,HO}&=-\dfrac{\hbar^2}{2(m+m_\text{D})}\partial_{X}^2+\dfrac{(m+m_\text{D})\omega^2X^2}{2}-\dfrac{\hbar^2}{2\mu_\text{FD}}\partial_{x_\text{rel}}^2+\dfrac{\mu_\text{FD}\omega^2x_\text{rel}^2}{2}+w(x_\text{rel})
\end{split}\\
&=H^\text{freeHO}(m+m_\text{D})+H^\text{HO}(\mu_\text{FD})
\end{align}
with reduced mass $\mu_\text{FD}=2m/3$, harmonic oscillator frequency $\omega$; the effective interaction $w$ and coordinates $x_\text{rel}=y-z$, $X=\frac{m_\text{D}y+mz}{m_\text{D}+m}$.
The center-of-mass (COM) wave function for fermion-dimer (FD) is just the common free harmonic oscillator (HO) solution of $H^\text{freeHO}(M)$ \cite{0cschwabl2007quantenmechanik}
\begin{align}
    \Psi_{n_\text{COM}}^\text{freeHO}(X)=\dfrac{1}{\sqrt{2^{n_\text{COM}} n_\text{COM}!}}\left(\dfrac{1}{l_\text{COM}^2\pi}\right)^{\dfrac{1}{4}}e^{-\dfrac{X^2}{2l_\text{COM}^2}}H_{n_\text{COM}}\left(\dfrac{X}{l_\text{COM}}\right)\text{ for } n_\text{COM}\in\mathbb{N}\label{eq: free Ho wave function}
\end{align}
with oscillator length $l_\text{COM}=\sqrt{\hbar/(3m\omega)}$, 
Hermite polynomials $H_{n_\text{COM}}$ and energy $E_{n_\text{COM}}=\hbar\omega(n_\text{COM}+\frac{1}{2})\ $. For the fermion-dimer system, the relative part of the Schr{\"o}dinger equation is solved by~(see, e.g,~\cite{Avakian1987,63busch1998two,61d2014three,64dehkharghani2016impenetrable}) 
\begin{align}
\label{eq: BuschDelta}
    \Psi^\text{HO,ET}_{\nu_\text{e}}(x_\text{rel})&\propto U\left(-\nu_\text{e},\dfrac{1}{2},\dfrac{x_\text{rel}^2}{l_\text{FD}^2}\right)e^{-\dfrac{x_\text{rel}^2}{2l_\text{FD}^2}}\\
    \begin{split}
\label{eq: BuschDeltaPrime1}
    \Psi^\text{HO,ET}_{\nu_\text{o}}(x_\text{rel})&\propto \dfrac{x_\text{rel}}{l_\text{FD}}U\left(-\nu_\text{o}+\dfrac{1}{2},\dfrac{3}{2},\dfrac{x_\text{rel}^2}{l_\text{FD}^2}\right)e^{-\dfrac{x_\text{rel}^2}{2l_\text{FD}^2}}\\&=\text{sign}(x_\text{rel})\Psi^\text{HO,ET}_{\nu_\text{e}=\nu_\text{o}}(x_\text{rel})
    \end{split}
\end{align}
with Tricomi's function $U(a,b,c)$, oscillator length $l_\text{FD}=\sqrt{\hbar/(\mu_\text{FD}\omega)}$ and quantum numbers $\nu_\text{e}$ and $\nu_\text{o}$. The quantum numbers of relative motion $\nu_j$ can be simply determined by numerically solving \cref{eq:HOEFTbc} of the main text, which follows from the boundary conditions \cref{eq:even channel eft+bc,eq:odd channel eft+bc}. 

To gain analytical insight into the properties of the system in the limit of strong attractive interactions, we expand $\nu_j$ in $1/\gamma^\text{HO}\approx0$ and solve \cref{eq:HOEFTbc} of the main text analytically. In the limit $\gamma^\text{HO}\to-\infty$, one finds the roots~\footnote{As discussed in the main text, we omit the (unphysical) deeply bound states.} 
\begin{align}
    \nu_\text{e}=\dfrac{1}{2},\ \dfrac{3}{2},\ \dfrac{5}{2},\ \dots~\text{ or }~\nu_\text{o}=0,\ 1,\ 2,\ \dots\ .
\end{align}
In general, one can derive the values of $\nu_j$ as an expansion in $1/\gamma^\text{HO}$ order by order analytically. Note that even and odd solutions are connected via the last equality in \cref{eq: BuschDeltaPrime1} so that the even  $\Psi^\text{HO,ET}_{\nu_\text{e}}$ and the odd $ \Psi^\text{HO,ET}_{\nu_\text{o}}$ solutions have the same energy $E^\text{HO,ET}_\text{rel}=\hbar\omega(2\nu_j+\frac{1}{2})$~\footnote{Up to the values of the quantum numbers $\nu_\text{e}$ and $\nu_\text{o}$, the only difference between even and odd solution is a sign switch at $x_\text{rel}=0$.}. With this the effective three-body energy reads
\begin{align}
E^\text{HO}_\text{FD}=E^\text{HO,ET}_\text{rel}+E_{n_\text{COM}}=\hbar\omega\left(2\nu_j+n_\text{COM}+1\right)
\end{align}
with free harmonic oscillator energy $E_n$. This can be compared with the overall three-body energy of the GY model by adding the exact dimer binding energy $E^\text{GY,HO}_\text{3-body}+B_\text{D}$.

{\it Dimer-dimer system.} Here, we consider the effective system of two dimers  ($N_\downarrow=2,N_\uparrow=2$).  We follow the same steps as for the even FD channel. First, we re-write the effective Hamiltonian as a free harmonic oscillator Hamiltonian for the center-of-mass part with total mass $4m$ plus a relative motion part with a $\delta$-interaction and reduced mass $\mu_\text{DD}=m$. The corresponding coordinates are $x_\text{rel}=y_1-y_2$, $X=\frac{m_\text{D}y_1+m_\text{D}y_2}{2m_\text{D}}$.
The free harmonic oscillator solution  is given by \cref{eq: free Ho wave function} with $l_\text{COM}=\sqrt{\hbar/(4m\omega)}$, so that we only have to investigate the relative motion part with the oscillator length $l_\text{DD}=\sqrt{\hbar/(\mu_\text{DD}\omega)}=l$. The corresponding  wave function reads
\begin{align}
    \Psi^\text{HO,ET}_{\nu_\text{e}}(x_\text{rel})&\propto U\left(-\nu_\text{e},\dfrac{1}{2},\dfrac{x_\text{rel}^2}{l_\text{DD}^2}\right)e^{-\dfrac{x_\text{rel}^2}{2l_\text{DD}^2}}
\end{align}
with quantum number $\nu_\text{e}$, which (analogous to FD case) is determined by
\begin{align}
\dfrac{1}{\gamma^\text{HO}}&\stackrel{!}{=}-\dfrac{\Gamma\left(-\nu_\text{e}\right)}{2\Gamma\left(-\nu_\text{e}+\dfrac{1}{2}\right)}\ .\label{eq:DDHOEFTbc}
\end{align}
The total wave function reads $\Psi^\text{DD}_{(n_\text{COM},\nu_\text{e})}(x_\text{COM},x_\text{rel})= \Psi_{n_\text{COM}}^\text{freeHO}(x_\text{COM})\Psi^\text{HO,ET}_{\nu_\text{e}}(x_\text{rel})$. This effective wave function enables analytical insights into the 4-body system in a harmonic oscillator close the the strong attractive limit. The quantum number $\nu_\text{e}$ can be used to calculate the effective dimer-dimer energy of the two dimer system as 
\begin{align}
E^\text{HO}_\text{DD}&=E^\text{HO,ET}_\text{rel}+E_{n_\text{COM}}=\hbar\omega\left(2\nu_\text{e}+n_\text{COM}+1\right)\ .
\end{align}
\begin{figure}
    \centering
    \includegraphics[width=0.48
    \textwidth]{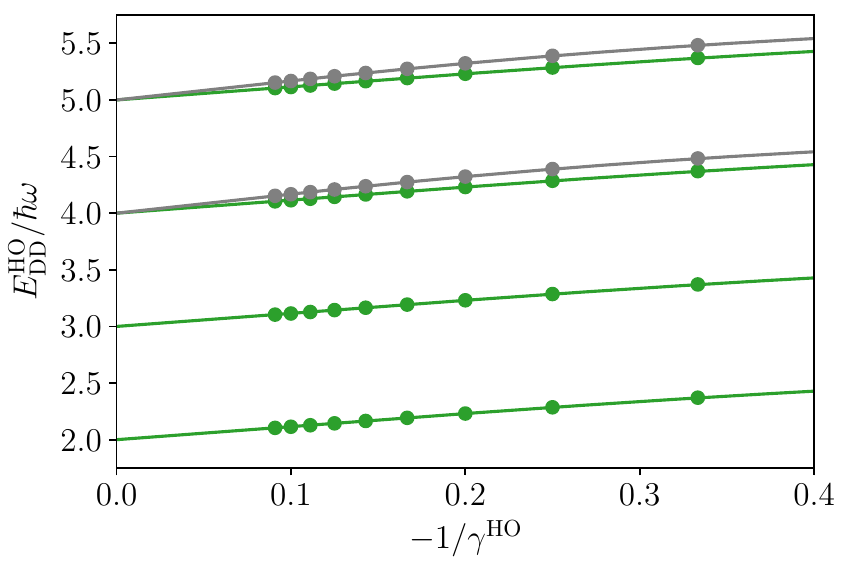}
    \caption{Energy of the effective dimer-dimer system in a harmonic trap. Solid curves show the results of \cref{eq:DDHOEFTbc}. Markers are numerical transcorrelated method data. Note that only even states occur in the dimer-dimer system. Identical colors indicate the same relative motion state.}
    \label{fig: DDHOFDTCMenergie}
\end{figure}
To compare the energy of the effective DD system with the energy of the original four-body system, one should add the binding energy for each dimer to the four-body energy,
$E^\text{GY,HO}_\text{4-body}+2B_\text{D}$, see \cref{fig: DDHOFDTCMenergie}. We observe a good agreement between our effective theory and the transcorrelated method for the considered values of $\gamma^\text{HO}$. 

\section{Boundary conditions as limit of a finite square well}\label{appendix:fsw}
Here, we investigate fermion-dimer scattering in relative coordinates assuming that particles are interacting via a square well (SW) potential. We focus on the limit $|kr_\text{int}|\ll1$ where the wave function inside the well is not resolved. In this limit we show the correspondence between the square well and the boundary conditions employed as the effective FD-interactions in the main text,  see \cref{eq:even channel eft+bc,eq:odd channel eft+bc}. With this we illustrate that one can interpret these boundary conditions as a limit of a square well interaction, at least for $|kr_\text{int}|\to0\ $. We remark that the limiting procedure will be carried out for even and odd channel independently -- one cannot describe both channels (even and odd), as they emerge in the GY model, with a single square well.

We consider an attractive square well potential $V^\text{SW}(|x_\text{rel}|\leq r_\text{int})=-V_0$ with interaction range $r_\text{int}$. Assuming the momentum $k$ outside the potential, the momentum inside reads $k_0=\sqrt{k^2+\frac{2\mu_\text{FD}}{\hbar^2}V_0}\ $. The corresponding odd and even wave functions read
\begin{align}
    \Psi^\text{SW}_\text{e}&\propto\begin{cases}\text{I:}\phantom{II} A_\text{e}\cos{(kx_\text{rel})}-B_\text{e}\sin{(kx_\text{rel})} \\
    \text{II:}\phantom{I}   \cos{(k_0x_\text{rel})}\\
    \text{III: }   A_\text{e}\cos{(kx_\text{rel})}+B_\text{e}\sin{(kx_\text{rel})}
    \end{cases}\label{eq:evenFSW wave function}\\
    \Psi^\text{SW}_\text{o}&\propto\begin{cases}\text{I:} -A_\text{o}\cos{(kx_\text{rel})}+B_\text{o}\sin{(kx_\text{rel})} \\
    \text{II:}\phantom{I} \sin{(k_0x_\text{rel})}\\
    \text{III: } A_\text{o}\cos{(kx_\text{rel})}+B_\text{o}\sin{(kx_\text{rel})}\ ,
    \label{eq:oddFSW wave function}
    \end{cases}
\end{align}
with sectors I: $x_\text{rel}<-r_\text{int}$, II: $|x_\text{rel}|\leq r_\text{int}$ and III: $x_\text{rel}>r_\text{int}$. To fix the parameters $A_j,B_j$, we use  continuity of the wave function and its derivative at $x_\text{rel}=\pm r_\text{int}$ (cf. Ref. \cite{0cschwabl2007quantenmechanik}).

Now, let us consider the limit $|kr_\text{int}|\ll1$. In this limit, the sector II (inside the well) is not resolved anymore, so that the square well potential resembles a point-like interaction. One can connect the wave function parts outside the square well potential via the interaction boundary conditions from \cref{eq:even channel eft+bc,eq:odd channel eft+bc} of the main text
\begin{align}
\partial_{x_\text{rel}}\Psi^\text{SW}_\text{e}\bigl|^{x_\text{rel}=r_\text{int}^+}_{x_\text{rel}=r_\text{int}^-}&\stackrel{|kr_\text{int}|\ll1}{=}\dfrac{g\mu_\text{FD}}{\hbar^2} \Psi^\text{SW}_\text{e}(\pm r_\text{int})\label{eq:deltaprime1fsw0}\\
\partial_{x_\text{rel}}^{2}\Psi^\text{SW}_\text{o}\bigl|^{x_\text{rel}=r_\text{int}^+}_{x_\text{rel}=r_\text{int}^-}&\stackrel{|kr_\text{int}|\ll1}{=}\dfrac{g\mu_\text{FD}}{\hbar^2}\partial_{x_\text{rel}}  \Psi^{\text{SW}}_\text{o}(\pm r_\text{int})\ ,\label{eq:deltaprime1fsw}
\end{align}
where $r_\text{int}^+$ means the limit from above $x_\text{rel}\searrow r_\text{int}$ and $r_\text{int}^-$ means the limit from below $x_\text{rel}\nearrow -r_\text{int}$.

{\it Even channel.} Let us first consider the square well potential and the boundary condition for the conceptually simpler even channel. Inserting the even-wave-function solution \cref{eq:evenFSW wave function} into the boundary condition \cref{eq:deltaprime1fsw0}, we end up with the expression
\begin{align}
    -2\lim_{kr_\text{int}\to 0}k_0\tan{(k_0r_\text{int})}=\dfrac{g\mu_\text{FD}}{\hbar^2}\Longrightarrow \dfrac{2}{3}\tilde{g}\stackrel{|kr_\text{int}|\ll1}{\approx}-2\sqrt{2\mu_\text{FD} V_0/\hbar^2}\tan{\left(\sqrt{2\mu_\text{FD} V_0/\hbar^2}\ r_\text{int}\right)}\ .
\end{align}
This condition \cref{eq:deltaprime1fsw0} in the limit $r_\text{int}\to 0$ leads to the scattering length \cite{22barlette2000quantum}
\begin{align}
    a_\text{e}=r_\text{int}+\dfrac{1}{\sqrt{2\mu_\text{FD} V_0/\hbar^2}\tan\left(\sqrt{2\mu_\text{FD} V_0/\hbar^2}\ r_\text{int}\right)}\to3r_\text{D}
\end{align}
and in the limit $1/\tilde{g}\to 0$ also to the effective range $r_\text{e}\to 0$. These are the values that appear for the FD-interaction within the GY model, see the main text.

{\it Odd channel.} Let us now consider the odd channel interaction. Making  use of the continuity of the wave function and its derivative in \cref{eq:deltaprime1fsw}, we arrive at
\begin{align}
    \lim_{kr_\text{int}\to 0}-2\dfrac{k^2}{k_0}\tan{(k_0r_\text{int})}&=\dfrac{g\mu_\text{FD}}{\hbar^2}\ .\label{eq:fsw22}
\end{align}
The limit in \cref{eq:fsw22} has to be carried out with the constraint $a_\text{o}=0\ $. Only then one can match the square well interaction with the odd channel boundary condition \cref{eq:deltaprime1fsw}. For a system at the threshold for binding of the $n\in\mathbb{N}$ bound state, $\sqrt{2\mu_\text{FD} V_0/\hbar^2}\ r_\text{int}\approx(n+1/2)\pi\ $~\cite{22barlette2000quantum}. We are interested in the case with $n=0$, which implies ($|kr_\text{int}|\ll1$) 
\begin{align}
    \tan(k_0r_\text{int})=&\pm\dfrac{2\sqrt{2\mu_\text{FD} V_0/\hbar^2}\ r_\text{int}}{k^2r_\text{int}^2}+\mathcal{O}\left(k^0r_\text{int}^0\right)\ .
\end{align}
The $\pm$ sign depends on whether we take the constraint of being at threshold from below $\nearrow(n+1/2)\pi$ or above $\searrow(n+1/2)\pi$. As there should be no bound trimer $n=0$, we consider the limit from below (+) and rewrite the boundary condition \cref{eq:deltaprime1fsw0,eq:deltaprime1fsw} as follows (here $\mu_\text{FD}=2m/3$)
\begin{align}
    \dfrac{2}{3}\tilde{g}&\stackrel{|kr_\text{int}|\ll1}{\approx}-\dfrac{4}{r_\text{int}}\ .\label{eq:fsw2}
\end{align}
At the threshold for the appearence of the first bound state $\sqrt{2\mu_\text{FD} V_0/\hbar^2}\ r_\text{int}\nearrow\pi/2$, the scattering length vanishes, $a_\text{o}=0$. It also holds that $r_\text{o}\to r_\text{int}$ (see Ref. \cite{22barlette2000quantum}) so that we end up with $r_\text{o}\to6r_\text{D}$.

To test the performance of the effective square well potential, we consider the problem on a ring, which is Bethe-ansatz solvable. For simplicity, we focus on the odd solution, where  $r_\text{o}=r_\text{int}\stackrel{!}{=}-6L/\gamma$ and $a_\text{o}=0$ (implying that $k_0r_\text{int}\nearrow\sqrt{k^2r_\text{int}^2+\pi^2/4}$). We impose the periodic boundary condition at $x_\text{rel}=\pm L/2$ as well as the continuity of the wave function \cref{eq:oddFSW wave function} and its derivative at $x_\text{rel}=\pm r_\text{int}$ leading to (assuming that $r_\text{int}<L/2\Longleftrightarrow \gamma<-12$):
\begin{align}
    \begin{split}
        \tan(kL/2)&=-A_\text{o}/B_\text{o}\\
        \sin(k_0r_\text{int})&=A_\text{o}\cos(kr_\text{int})+B_\text{o}\sin(kr_\text{int})\\
        k_0\cos(k_0r_\text{int})&=k\left(-A_\text{o}\sin(kr_\text{int})+B_\text{o}\cos(kr_\text{int})\right)\ .
    \end{split}
\end{align}
We eliminate the coefficients $A_\text{o}$ and $B_\text{o}$ from this equation, which leads to the equation on the relative momentum $k$
\begin{align}
    \dfrac{k_0\cot(k_0r_\text{int})\left(\tan(kr_\text{int})\cot(kL/2)-1\right)}{k\left(\cot(kL/2)+\tan(kr_\text{int})\right)}=1\ .\label{eq:FSW ring energy fix}
\end{align}
Once this equation is solved, the energy of the FD system can be calculated as follows: $E_\text{FD}=3\hbar^2k^2/4m$.

 To compare the exact energies from the GY model with those calculated using the square well interaction and the boundary condition from \cref{eq:odd channel eft+bc} (see \cref{fig: FDEFT vs GY model} in the Letter), we assume the following expansion of the ground state energy for large values of t $1/\gamma=0$ as
\begin{equation}
    E_\text{FD}\dfrac{mL^2}{\hbar^2}=\dfrac{3\pi^2}{4}\left(1+\dfrac{c_1}{\gamma}+\dfrac{c_2}{\gamma^2}+\dfrac{c_3}{\gamma^3}+\mathcal{O}\left(\gamma^{-4}\right)\right)\ .
\end{equation}
The extracted parameters for the GY model can then be compared with the parameters for square well- and boundary condition interaction, which can be determined analytically from \cref{eq:deltaprime ring energy fix,eq:FSW ring energy fix}. The resulting parameters coincide for all three theories up to the second order in $1/\gamma$: $c_1\approx-12$ and $c_2\approx108$. At the third order, the parameters deviate $c^{\text{GY}}_3\approx-404$, $c_3^\text{SW}\approx-787$, and $c_3^\text{ET}\approx-509$. We remark that the parameter $c^{\text{GY}}_3$ agrees better with  $c_3^\text{ET}$ than with $c_3^\text{SW}$, see also the inset of \cref{fig: FDEFT vs GY model} in the main text.

{\it Peculiarities of the odd-channel problem.}  
The step from the square-well potential to the boundary condition in Eq.~(\ref{eq:deltaprime1fsw}) is fully justified only in the limit $1/\tilde{g}=0$, i.e., when the width of the square well is vanishing. Otherwise, this step neglects the part of the wave function inside the well. This leads to the broken causality and non-hermicity (violated Wigner lower limit) of the problem defined via the boundary condition. To directly show this, one can just demonstrate non-orthogonality of the eigenstates $\Psi_i$ and $\Psi_j$ with $E_i\neq E_j$, on a ring or in an harmonic oscillator (see \cref{eq:odd sol ring,eq: BuschDeltaPrime1}). Indeed, the scalar product
is non-vanishing, $\braket{\Psi_i|\Psi_j}\neq 0$, which demonstrates that the problem is indeed non-Hermitian.

\section{Square well potential for a fermion-dimer system in a harmonic trap}\label{S5ESWHO}
Here, we use a square well (SW) potential to describe the effective interaction in the odd channel for a trapped FD system, i.e., we use the potential from the previous section, but now with a harmonic confinement. As it appears impossible to describe the even and odd  channels for a FD system with a single square well potential, we only treat here the odd channel -- the even channel allows for a simple description with a $\delta$-function and in principle does not require any further conceptual developments. We write the corresponding wave function as
\begin{align}
     &\Psi^\text{HO,SW}_{\nu_\text{o}}\propto\begin{cases}
         \text{$|x_\text{rel}|\geq r_\text{int}$: }A_\text{HO}\Psi^\text{HO,ET}_{\nu_\text{o}}(x_\text{rel})\\
         \text{$|x_\text{rel}|< r_\text{int}$: } \sin(kx_\text{rel})
         \end{cases}\label{eq:SW HO approx sol}\\
         \begin{split}
        \text{with } \quad  &A_\text{HO}=\sin(kr_\text{int})/\Psi^\text{HO,ET}_{\nu_\text{o}}(r_\text{int})  \quad \text{and} \quad kl=\sqrt{\dfrac{4}{3}\left(2\nu_\text{o}+\dfrac{1}{2}\right)+\dfrac{\pi^2{\gamma^\text{HO}}^2}{144}}.
         \label{eq:continuity FSW HO}
         \end{split}
\end{align}
We are neglecting the effect of the harmonic trap in Eq.~(\ref{eq:SW HO approx sol}), i.e., we assume that $v_\text{ext}\approx0$ if $|x_\text{rel}|< r_\text{int}$. This approximation is valid if $|l/r_\text{int}|\ll1\Leftrightarrow-1/\gamma^\text{HO}\ll1/6$. Equation~(\ref{eq:continuity FSW HO}) comes from the continuity of the wave function and the comparison of the energy in- and outside the square well 
\begin{align}
    \dfrac{\hbar^2k^2}{2\mu_\text{FD}}-V_0=\hbar\omega(2\nu_\text{o}+\dfrac{1}{2})\ .
\end{align}
The continuity of the wave function's derivative fixes the quantum number $\nu_\text{o}$
\begin{align}
    k\cos(kr_\text{int})\stackrel{!}{=}A_\text{HO}\partial_{x_\text{rel}}\Psi^\text{HO,ET}_{\nu_\text{o}}(x_\text{rel}=r_\text{int})\ ,\label{eq:SW nu cond}
\end{align}
and thereby the total energy of the effective fermion-dimer system (compare with \cref{eq:EHOFD} of the main text)
\begin{align}
E^\text{HO}_\text{FD}=\hbar\omega\left(2\nu_j+n_\text{COM}+1\right)\ . 
\end{align}

\begin{figure}
    \centering
    \includegraphics[width=0.48
    \textwidth]{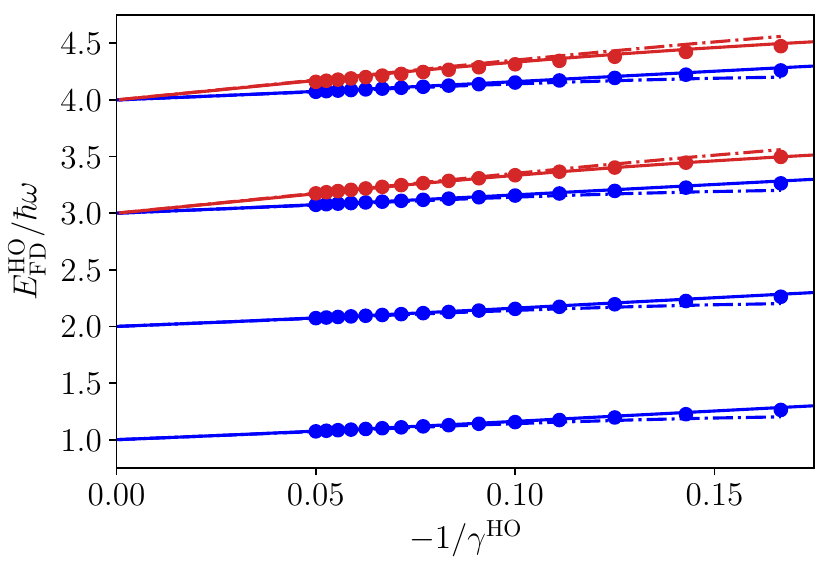}
    \caption{Energy of the odd channel of effective fermion-dimer system in a harmonic trap. Solid curves show the odd channel boundary condition result of \cref{eq:HOEFTbc}. Dash-dotted curves represent the approximate square well solution results of \cref{eq:SW nu cond}. Markers are numerical transcorrelated method data, serving as reference value to compare the boundary condition against the approximate square well result. The
    approximate square well solution is only valid within $-1/\gamma^\text{HO}\ll1/6$. Identical colors indicate the same relative motion state.}
    \label{fig:HOFSW}
\end{figure}

In \cref{fig:HOFSW} we compare the energies calculated using the boundary conditions from~\cref{eq:odd channel eft+bc} of the main text against $E^\text{HO}_\text{FD}$. We see that both methods agree in the limit of strong interactions. Further, by comparing our results to the numerical calculations, we observe that the boundary-condition description of the fermion-dimer system yields energies that are more accurate than those produced by the square well.  This observation agrees with our discussion of a fermion-dimer system in a ring (see \cref{fig: FDEFT vs GY model} of the main text).

In summary, the square well potential can be used as a finite-range effective interaction. However, it leads to a model that is harder to solve compared to the boundary condition formulation. Further, it leads to the energy spectrum that appears less accurate. 
The only clear advantage of the SW potential is that is satisfies the Wigner lower limit. Otherwise, the boundary conditions presented in the main text simplify calculations and provide a more comprehensible physical picture of the problem.

\section{Fermi gas with an impurity}\label{S6Manybody}
\label{appendix: PolaronEnergyManyFermions}

Here, we explain how we calculate the impurity energy for a system with one spin-down and $N_\uparrow$ spin-up fermions, see Fig.~\ref{fig: ThermoLimit} of the main text. As we consider strong interactions, we use the effective description and reformulate the problem in terms of  $M=N_\uparrow-1$ spin polarized fermions and one dimer interacting via effective interactions. To calculate the energy of this system, we follow the prescription given in Ref.~\cite{Wlodzynski2022}: We perform a generalized transformation to the relative and center-of-mass frame (somewhat similar to the coordinate space `Lee-Low-Pines' transformation~\cite{Lee1953}) which results in two decoupled Hamiltonians:
\begin{align}
    H^\text{cm}&=-\dfrac{\hbar^2}{2(N_\uparrow+1)m}\partial_y^2+\dfrac{(N_\uparrow+1)m\omega^2}{2}y^2\\
    H^\text{rel}&=-\dfrac{\hbar^2}{2\mu_\text{FD}}\sum\limits_{i=1}^{N_\uparrow-1}\partial_{z_i}^2+\dfrac{m\omega^2}{2}\dfrac{N_\uparrow}{N_\uparrow+1}\sum\limits_{i=1}^{N_\uparrow-1} z_i^2+\sum\limits_{i=1}^{N_\uparrow-1}w(z_i)+\sum\limits_{i, j}V(z_i, z_j)
    \label{eq: Hamiltonian LLP}
\end{align}
with $V(z_i, z_j)=-\dfrac{\hbar^2}{4m}\partial_{z_i}\partial_{z_j}-\dfrac{m\omega^2}{2(N_\uparrow+1)}z_iz_j$ and $w$ the effective interaction for the fermion-dimer scattering discussed in the main text. For the even channel we employ a contact interaction, 
\begin{equation}
w(x_\text{rel})=\frac{g}{2}\delta(x_\text{rel})\,,    
\end{equation}
while in the odd channel, we use a finite square well interaction
\begin{equation}
    w(x_\text{rel})=\begin{cases}
    -\frac{\pi^2\hbar^2}{8\mu_{\mathrm{FD}}(r_{\mathrm{o}}^{\mathrm{FD}})^2}&\text{ if } |x_\text{rel}|<r_{\mathrm{o}}^{\mathrm{FD}},\\
    0 &\text{ else.}
    \end{cases}
\end{equation}

The center-of-mass Hamiltonian can be solved analytically; the eigenstates are the ones of the harmonic oscillator. For the relative part, we employ the configuration interaction method~\cite{CremonPhDThesis, BjerlinPhdThesis}. We use a one-body basis which solves the one-body part of the relative Hamiltonian (including the effective theory interaction). The solutions for the even and the odd-channel interaction are given by~\cref{eq: BuschDelta} and~\cref{eq:SW HO approx sol}. Next, we employ the formalism of second quantization in which the Hamiltonian reads
\begin{equation}
    H=\sum\limits_{ij}A_{ij} a_i^\dagger a_j + \sum\limits_{ijkl} B_{ijkl} a_i^\dagger a_j^\dagger a_k a_l\label{eq:manybody hamiltonian}
\end{equation}
with $a_i^\dagger$ ($a_i$) fermionic creation (annihilation) operators. The matrices $A_{ij}$ and $B_{ijkl}$ are one- and two-body matrices whose elements are calculated using the one-body basis. We construct antisymmetric $M=N_\uparrow-1$ particle basis states and then, we build the Hamiltonian matrix that is finally diagonalized using the Arnoldi/Lanczos algorithm~\cite{golubmatrix}.
We truncate the Hamiltonian matrix by introducing a one-body basis cutoff, $n$. 

\begin{figure}
    \centering
    \includegraphics[width=0.5
    \textwidth]{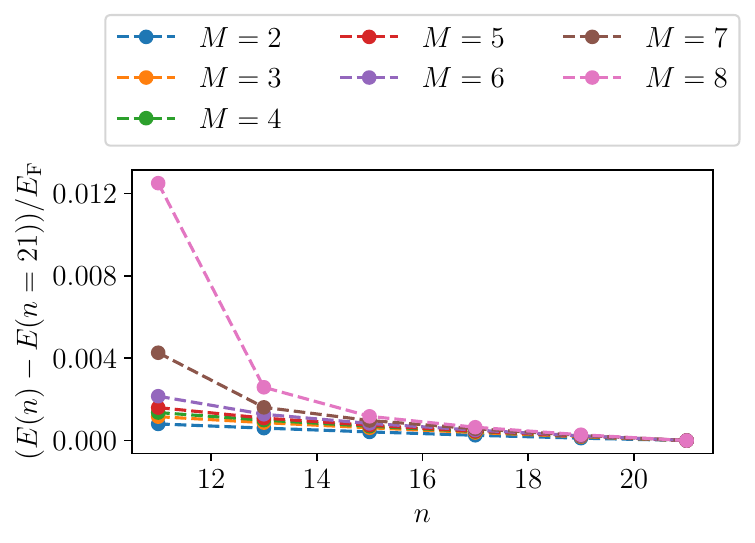}
    \caption{Convergence plot of the energy for different numbers of unpaired fermions $M=N_\uparrow-1$  obtained by diagonalizing Eq.~\eqref{eq: Hamiltonian LLP}. The strength of the interaction is fixed at $\gamma_\text{scaled}=-10$. We plot the energy difference for a one-body cutoff $n$ with the highest cutoff used in our calculation, $n=21$, as a function of $n$. To compare the different particle numbers, we scale the energy difference by the Fermi energy.}
    \label{fig:LLP Convergence}
\end{figure}

We show in Fig.~\ref{fig:LLP Convergence} a convergence plot for an interaction strength of $\gamma_\text{scaled}=-10$, which is the lowest interaction strength in the main text, and hence its wave function is the most correlated (recall that the strongly interacting problem can be mapped onto a weakly interacting one). We note that larger systems require more one-body basis states for convergence. The reason for this is that less orbitals are unoccupied. We found that a number of $n=21$ basis states is sufficient to reach converged results.

\section{The transcorrelated method}\label{S7TCM}

To serve as a benchmark for the effective theory in a harmonic oscillator, we employ numerical exact diagonalization with the transcorrelated method for particles with contact interactions~\cite{68jeszenszki2018accelerating,Jeszenszki2020}. The transcorrelated method~\cite{Boys1969} starts by folding a Jastrow factor $e^\tau$ into the Hamiltonian $H$ of Eq.~(1) in the main text via a similarity transformation
\begin{equation}\label{eq:tc-similarity}
  \tilde{H} = e^{-\tau} H e^{\tau}\,,
\end{equation}
where
\begin{equation}\label{eq:tc-jastrow}
  \tau = \sum_{i<j}^{N} u(x_i - x_j)\,,
\end{equation}
is a Jastrow factor~\cite{Jastrow1955}, $x_i$ and $x_j$ are positions of the $i$-th and $j$-th particles in the many-body state, and the function $u$ is designed to describe the cusp that occurs in the wave function due to the $\delta$-function interaction~\cite{68jeszenszki2018accelerating}. Specifically, it is defined through its Fourier transform
\begin{equation}\label{eq:tc-fourier}
  u(x) = {(2\pi)}^{-1}\int \exp(-ikx)\tilde{u}(k)\mathrm{d}k ,
\end{equation}
where
\begin{equation}\label{eq:tc-correlation}
  \tilde{u}(k) = \begin{cases}
                   -\tilde{g} / k^2 & \text{if $|k| \ge k_c$,} \\
                   0 & \text{otherwise,}
                 \end{cases}
\end{equation}
$\tilde{g}=mg/\hbar^2$ is the inverse dimer size, and $k_c$ is a small-momentum cutoff parameter. The value of $k_c$ determines the shape of the Jastrow factor at large distances. For a sufficiently large size of the single-particle basis the results become insensitive to the value of $k_c$.

The similarity transformation of Eq.~\eqref{eq:tc-similarity} does not affect the external potential or the contact interaction in the Hamiltonian, as these terms are diagonal in real space. The kinetic energy operator generates additional terms including three-body interactions, which we fully account for.
The resulting transcorrelated Hamiltonian $\tilde{H}$ is non-Hermitian, but its low-lying eigenvalues converge much faster to the infinite basis set limit when the operator is expanded in a truncated single-particle basis with $n$ plane waves. In this case, the convergence improves from $n^{-1}$ to $n^{-3}$. For a detailed description of the method, see Ref.~\cite{68jeszenszki2018accelerating}.

\begin{figure}
    \includegraphics[width=.8\textwidth]{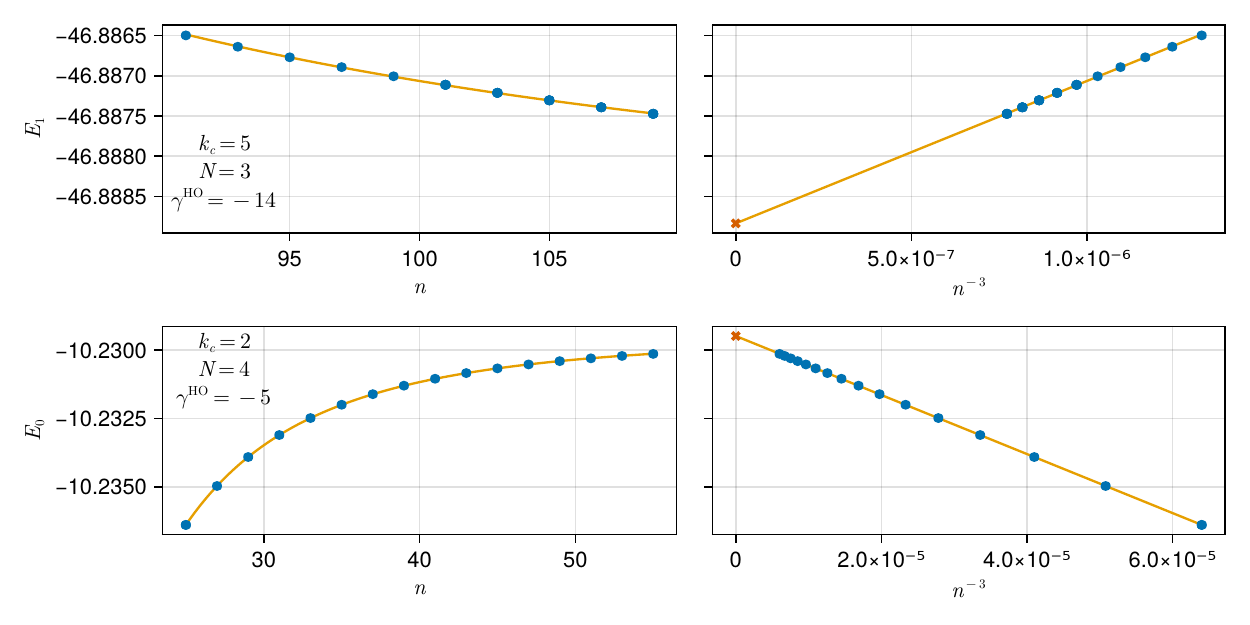}
    \caption{\label{fig:tc-convergence} 
    Convergence of the energy and extrapolation to the infinite basis set limit for the transcorrelated method. The energy is plotted as a function of the number $n$ of single-particle basis functions on different scales (left and right). The top and bottom panels show data for two representative sets of parameters and particle numbers as indicated in the plot. The blue data points are numerical results using the transcorrelated method and the orange line is a fit of $E(n) = E +\alpha n^{-3}$, where $E$ and $\alpha$ are fitted parameters. The red crosses on the right hand panes mark the extrapolated values $E$ of the infinite basis set limit $n\to\infty$, which is the data reported in the main text (Fig.~\ref{fig: HOFDTCMenergie}). A constant value of $k_c$ is used in the extrapolation procedure.
    }
\end{figure}

We formulate the transcorrelated Hamiltonian $\tilde{H}$ including the harmonic trapping potential in momentum space on a ring of length $L= 7l$ (where $l=\sqrt{\hbar/(m\omega)}$ is the harmonic oscillator length) and work with a finite number of momentum modes $n$.
For each value of $n$, we first realize the operator $\tilde{H}$ as a matrix $\mathbf{H}$ in a Fock space basis. Then, we set $k_c$ by minimizing the variance of $\mathbf{H}$ with respect to a single Fock state and use the standard Arnoldi method to find its eigenvalues.
All calclulations were done using the open-source package \texttt{Rimu.jl} \cite{rimucode}.
We repeat the process for various momentum cutoffs $n$ and extrapolate the data to the infinite basis set limit by fitting $E(n) = E + \alpha n^{-3}$ to the data and reporting $E$, as shown in Fig.~\ref{fig:tc-convergence}. 

This procedure works well across a wide range of interaction strengths, however, for a larger number of particles or stronger interactions, a larger number of momentum modes is required. Data presented in Fig.~\ref{fig: HOFDTCMenergie} of the main text and in Figs.~\ref{fig: DDHOFDTCMenergie}, and~\ref{fig:HOFSW} of the supplement includes interaction strengths up to $\gamma^{\mathrm{HO}}=-20$ for $N=2$ and $3$ particles, and $\gamma^{\mathrm{HO}}=-11$ for $N=4$. When optimizing the cutoff parameter $k_c$, we have found that for a given $N$ and $\gamma^{\mathrm{HO}}$, it varies when $n$ is small, but stabilizes rather quickly. The optimal value of $k_c$ appears to be correlated with interaction strength --- it is higher for stronger interactions.
When performing the extrapolation, we use a range of $n$ on which the optimal value of $k_c$ is stable and the trend $E(n) - E \propto n^{-3}$ is followed.

\section{Fermi gas with an impurity: Many-body perturbation theory}\label{S8Manybodyperturbation}

To calculate the energy of an impurity in a Fermi gas, we need to calculate the expectation value (see the main text for details)
\begin{equation}
\langle \phi| V_{\text{FD}}|\phi\rangle = \sum_{n=0}^{N_{\uparrow}-1} \Delta E_2 (n),
\label{eq:app:VFD}
\end{equation}
where $\phi$ is a ground-state wave function of a system with $M=N_\uparrow-1$ spin polarized fermions and one dimer in the limit $1/g=0$;
$\Delta E_2(n)$ is a first-order perturbative contribution to the two-body energy assuming that the non-interacting two-body state places a fermion in the $n$th state of the harmonic oscillator. To compute $\Delta E_2 (n)$, we work in the relative coordinates
\begin{equation}
\Delta E_2(n)=\sum_{i=0}^{n}\alpha_{i}(n)^2\Delta E^{\text{rel}}_2(i),
\end{equation}
where $\alpha_i(n)$ is the  Talmi-Moshinsky-Smirnov coefficient~\cite{Talmi1952}:
\begin{equation}
\alpha_i(n)=\int \mathrm{d}y\mathrm{d}z \phi^{\text{D}}_0(y)\phi_n^{\text{F}}(z)\phi^{\text{COM}}_{n-i}(X)\phi_i^{\text{rel}}(x_\text{rel}).
\end{equation}
$\Delta E^{\text{rel}}_2(i)$ is the energy change in relative coordinates. It depends on the parity of $i$. If $i$ is even, then 
\begin{equation}
\Delta E_2^{\text{rel}}(i)=-\frac{\sqrt{6}}{ \gamma^\text{HO} \sqrt{\pi}}\frac{(2i+1)}{2^{i/2} (i/2)!}(i-1)!!;
\end{equation}
if $i$ is odd, then
\begin{equation}
\Delta E_2^{\text{rel}}(i)=-\sqrt{\frac{3}{2}}\frac{8}{ \gamma^\text{HO} \sqrt{\pi}}\frac{i !!}{2^{(i+1)/2} (\frac{i-1}{2})!}.
\end{equation}
These are the corrections for the two-body energies. 

Let us give an example of a computation of Eq.~(\ref{eq:app:VFD}) for 1+2: 
\begin{equation}
\Delta E_{1+2} = \Delta E_2 (0)+\Delta E_2 (1).
\end{equation}
The first term in this expression can be easily calculated: $\Delta E_2 (0)=-\sqrt{6}/(\gamma^\text{HO} \sqrt{\pi})$. The calculation of $\Delta E_2 (1)$ is somewhat more involved 
\begin{equation}
\Delta E_2(1) = \alpha_0(1)^2\Delta E_2^{\text{rel}}(0)+\alpha_1(1)^2 \Delta E_2^{\text{rel}}(1).
\end{equation}
By collecting all terms, we obtain
\begin{equation}
\Delta E_{1+2} = -\sqrt{\frac{3}{2}}\frac{16}{3\gamma^\text{HO} \sqrt{\pi}},
\end{equation}
For the 1+3 system, we have
\begin{equation}
\Delta E_{1+3} = \Delta E_{1+2} + \Delta E_2(2)=-\sqrt{\frac{3}{2}}\frac{86}{9\gamma^\text{HO} \sqrt{\pi}}.
\end{equation}
Further, $\Delta E_{1+4}=-\sqrt{\frac{3}{2}}\frac{392}{27\gamma^\text{HO} \sqrt{\pi}}, \; \Delta E_{1+5}=-\sqrt{\frac{3}{2}}\frac{1630}{81\gamma^\text{HO} \sqrt{\pi}},\; \Delta E_{1+6}=-\sqrt{\frac{3}{2}}\frac{6392}{243\gamma^\text{HO} \sqrt{\pi}}, \; \Delta E_{1+7}=-\sqrt{\frac{3}{2}}\frac{2674}{81\gamma^\text{HO} \sqrt{\pi}}, \; \Delta E_{1+8}=-\sqrt{\frac{3}{2}}\frac{89056}{2187\gamma^\text{HO} \sqrt{\pi}}$. These values were used in Fig. 3 of the main text.

\section{Fermi gas with an impurity: Thermodynamic limit}\label{S9}
In the thermodynamic limit, the effect of the trap should be negligible and the energy of the impurity can be calculated from the underlying GY model using the Bethe ansatz~\cite{8mcguire1966interacting}
\begin{equation}
\frac{E_{\mathrm{polaron}}}{E_F}=-\frac{2}{\pi}\left[\frac{|\gamma_\text{scaled}|}{2\pi }+\mathrm{tan}^{-1}\left(\frac{|\gamma_\text{scaled}|}{2\pi}\right)+\left(\frac{|\gamma_\text{scaled}|}{2\pi}\right)^2\left(\frac{\pi}{2}+\mathrm{tan}^{-1}\left(\frac{|\gamma_\text{scaled}|}{2\pi}\right)\right)\right].
\end{equation}
We use this expression to calculate the limiting values of the energies in Fig. 3. 
In the limit $|g|\to\infty$, we find 
\begin{equation}
\frac{E_{\mathrm{polaron}}}{E_F}\simeq -\frac{\gamma_\text{scaled}^2}{2\pi^2}-1+\frac{8}{3 |\gamma_\text{scaled}|}.
\end{equation}
The first term here describes the energy of the dimer. The second is the energy gain from removing one fermion from the Fermi level. The rest is the interaction energy discussed in the main text.  

\twocolumngrid
\bibliography{ref}

\end{document}